\documentclass[sigconf]{acmart}
\usepackage[portuguese]{babel}

\usepackage{graphicx}
\usepackage{textcomp}
\usepackage{verbatim}
\usepackage{booktabs}

\AtBeginDocument{%
  }

\acmConference[SBQS'25]{Brazilian Symposium on Software Quality}{Nov 04--07, 2025}{São José dos Campos, SP}
\setcopyright{none}
\renewcommand\footnotetextcopyrightpermission[1]{}

\begin{document}
\title{Lessons Learned from the Use of Generative AI in Engineering and Quality Assurance of a WEB System for Healthcare}

\author{Guilherme H. Travassos}
\email{ght@cos.ufrj.br}
\affiliation{%
\institution{Universidade Federal do Rio de Janeiro (UFRJ)}
\city{Rio de Janeiro, RJ}
\country{Brasil}}

\author{Sabrina Rocha}
\email{sabrinarocha@cos.ufrj.br}
\affiliation{%
 \institution{Universidade Federal do Rio de Janeiro (UFRJ)}
  \city{Rio de Janeiro, RJ}
  \country{Brasil}}
  
 \author{Rodrigo Feitosa}
 \email{rfeitosa@cos.ufrj.br}
 \affiliation{%
  \institution{Universidade Federal do Rio de Janeiro (UFRJ)}
   \city{Rio de Janeiro, RJ}
   \country{Brasil}}

 \author{Felipe Assis}
 \email{fassis@cos.ufrj.br}
 \affiliation{%
 \institution{Universidade Federal do Rio de Janeiro (UFRJ)}
 \city{Rio de Janeiro, RJ}
 \country{Brasil}}

  \author{Patrícia Gonçalves}
 \email{patriciaamaralgurgel@ufrj.br}
 \affiliation{%
 \institution{Universidade Federal do Rio de Janeiro (UFRJ)}
 \city{Rio de Janeiro, RJ}
 \country{Brasil}}

\author{André Gheventer}
\email{gheventer@ufrj.br}
\affiliation{%
 \institution{Universidade Federal do Rio de Janeiro (UFRJ)}
  \city{Rio de Janeiro, RJ}
  \country{Brasil}}
  
 \author{Larissa Galeno}
 \email{galeno@cos.ufrj.br}
 \affiliation{%
   \institution{Universidade Federal do Rio de Janeiro (UFRJ)}
   \city{Rio de Janeiro, RJ}
   \country{Brasil}}

 \author{Arthur Sasse}
 \email{artsasse@cos.ufrj.br}
 \affiliation{%
 \institution{Universidade Federal do Rio de Janeiro (UFRJ)}
 \city{Rio de Janeiro, RJ}
 \country{Brazil}}

\author{Júlio César Guimarães}
\email{jcguimaraes@cos.ufrj.br}
\affiliation{%
  \institution{Universidade Federal do Rio de Janeiro (UFRJ)}
  \city{Rio de Janeiro, RJ}
  \country{Brasil}}
  
 \author{Carlos Brito}
 \email{carloshenriquefbf@poli.ufrj.br}
 \affiliation{%
   \institution{Universidade Federal do Rio de Janeiro (UFRJ)}
   \city{Rio de Janeiro, RJ}
   \country{Brasil}}

 \author{João Pedro Wieland}
 \email{jpvbwieland@cos.ufrj.br}
 \affiliation{%
 \institution{Universidade Federal do Rio de Janeiro (UFRJ)}
 \city{Rio de Janeiro, RJ}
 \country{Brasil}}

\renewcommand{\shortauthors}{Travassos et. al}

\begin{abstract}
The advances and availability of technologies involving Generative Artificial Intelligence (AI) are evolving clearly and explicitly, driving immediate changes in various work activities. Software Engineering (SE) is no exception and stands to benefit from these new technologies, enhancing productivity and quality in its software development processes. However, although the use of Generative AI in SE practices is still in its early stages, considering the lack of conclusive results from ongoing research and the limited technological maturity, we have chosen to incorporate these technologies in the development of a web-based software system to be used in clinical trials by a thoracic diseases research group at our university. For this reason, we decided to share this experience report documenting our development team's learning journey in using Generative AI during the software development process. Project management, requirements specification, design, development, and quality assurance activities form the scope of observation. Although we do not yet have definitive technological evidence to evolve our development process significantly, the results obtained and the suggestions shared here represent valuable insights for software organizations seeking to innovate their development practices to achieve software quality with generative AI.
\end{abstract}


\maketitle

\section{Introdução}

O ciclo de desenvolvimento de software compreende fases estruturadas que orientam a concepção, implementação e manutenção de sistemas de software \cite{8047358}. Tradicionalmente, esse processo depende fortemente de atividades manuais, o que o torna propenso a erros e frequentes atrasos inesperados. Com o aumento da demanda por soluções com mais agilidade e escaláveis, torna-se necessário incorporar tecnologias capazes de acelerar diferentes fases do ciclo de desenvolvimento e a garantia da qualidade do software \cite{10894980}.

Os avanços em tecnologias de Inteligência Artificial (IA) generativa têm impulsionado transformações relevantes em diversos processos de trabalho, incluindo a Engenharia de Software (ES) \cite{modi2024transforming, Almeida2024}. Tecnologias baseadas em grandes modelos de linguagem (LLMs) e modelos generativos podem oferecer apoio para (semi) automatizar atividades ao longo do ciclo de desenvolvimento de software. Esse movimento vem alterando a expectativa de como sistemas de software podem ser concebidos, construídos e mantidos, visando obter ganhos em qualidade e produtividade, reduzir custos e garantir maior velocidade nas entregas \cite{sauvola2024future}. Apesar do avanço rápido na promoção dessas tecnologias, ainda existem dúvidas sobre sua disponibilidade, efetividade, limitações e melhores práticas quando aplicadas em projetos reais de desenvolvimento \cite{coutinho2024role}. 

Diante desse cenário de transformação, este relato de experiência descreve o desenvolvimento de um sistema de software WEB na área da saúde. O sistema foi criado para apoiar profissionais de saúde no acompanhamento clínico de pacientes em tratamento para cessação do tabagismo. A iniciativa foi conduzida por nossa equipe de desenvolvimento a partir de uma demanda do Instituto de Doenças do Tórax da Universidade Federal do Rio de Janeiro (IDT/UFRJ). Decidimos fazer uso de ferramentas de IA generativa para apoiar a construção do produto. Nosso interesse, além da efetiva entrega de uma solução de software com qualidade, inclui observar, na prática, como tecnologias emergentes, incluindo recursos de IA generativa, podem ser integradas ao processo de construção de sistemas de software com qualidade em contextos reais.

Como parte do processo de construção do software, realizamos o mapeamento de ferramentas baseadas em IA generativa com potencial para oferecer suporte às diferentes fases do ciclo de desenvolvimento de software \cite{rocha25metaprotocol}. Os resultados indicaram a existência de algumas poucas soluções industriais voltadas a atividades como levantamento de requisitos, projeto, gerenciamento de projetos, codificação, testes, experimentação e, de forma transversal, garantia da qualidade. No entanto, observamos que há poucas evidências empíricas sobre a eficácia dessas tecnologias quando aplicadas em projetos reais, apesar das inúmeras “afirmações” sobre sua viabilidade e capacidade. Entendemos que essa escassez de evidências representa uma oportunidade de aprendizado. Para isso, adaptamos nosso processo de desenvolvimento interativo e incremental para integrar, na maior medida possível, ferramentas baseadas em IA generativa ao longo de todo o ciclo de desenvolvimento. Nossa perspectiva de observação foi direcionada pela seguinte questão: \textit{``é possível desenvolver um sistema de software utilizando apenas ferramentas baseadas em IA generativa nas diferentes fases do ciclo de desenvolvimento?''}

Os resultados indicaram que, embora a aplicação exclusiva de IA generativa ao longo do ciclo de desenvolvimento ainda não permita a construção de um sistema WEB funcional, sua adoção contribuiu significativamente para a evolução dos processos construtivos. A experiência revelou avanços, mas também desafios, especialmente no que diz respeito à adaptação das práticas tradicionais de engenharia às novas dinâmicas introduzidas por essas tecnologias. Um dos principais achados foi o surgimento de novos artefatos no projeto, como os \textit{prompts}, que passaram a exigir atenção semelhante à dedicada a elementos clássicos, como requisitos e casos de teste.

Um \textit{prompt} pode ser definido como uma instrução ou informação fornecida a uma ferramenta de IA generativa, para gerar uma resposta, executar uma tarefa ou orientar o comportamento do modelo \cite{zhou2022large}. Quanto mais claro, específico e detalhado for o \textit{prompt}, maiores são as chances de se obter uma resposta útil, criativa e precisa — sendo aplicável em diversos contextos, como geração de textos, imagens, código, traduções, entre outros \cite{zhou2022large}. Na prática, observou-se que o \textit{prompt} tornou-se um elemento central na comunicação entre desenvolvedores e tecnologias de IA generativa. A qualidade das respostas obtidas mostrou-se fortemente dependente da clareza, completude e intencionalidade expressas na formulação desses \textit{prompts}, o que exigiu dos desenvolvedores habilidades relacionadas à escrita técnica e na definição precisa dos objetivos. 

Outro elemento de destaque foi o uso do \textit{Markdown}, uma linguagem de marcação, comumente utilizada para formatar textos de forma rápida e funcional, especialmente em ambientes digitais como blogs, fóruns, documentação técnica e plataformas de desenvolvimento, como o \textit{GitHub}. Com uma sintaxe simples, baseada em caracteres comuns (como asteriscos, \textit{hash} e colchetes), o \textit{Markdown} permite a criação de títulos, listas, negrito, itálico, \textit{links}, imagens e blocos de código, sem a necessidade de comandos complexos. Sua principal vantagem reside na clareza e na praticidade: mesmo sem renderização, o conteúdo permanece legível, facilitando tanto a edição quanto a colaboração entre desenvolvedores \cite{xie2018r}. 
 
Assim, ao longo do projeto, o \textit{Markdown} revelou-se essencial não apenas para organizar os resultados gerados pelos modelos, mas também para documentar \textit{prompts}, decisões de projeto e instruções técnicas. Esses artefatos passaram a desempenhar um papel relevante nas atividades de garantia da qualidade, sobretudo nas atividades de requisitos e testes, que exigiram maior esforço de inspeção e validação dos artefatos gerados com o apoio de IA generativa. Neste sentido, este relato de experiência descreve as ações e observações realizadas ao longo do processo construtivo, destacando o aprendizado sobre pontos fortes e fracos observados. 

O restante deste artigo está organizado da seguinte forma: a Seção 2 apresenta o caso de observação; a Seção 3 descreve o processo construtivo com o uso de IA generativa; a Seção 4 relata sua aplicação na engenharia de garantia da qualidade de um sistema de software; a Seção 5 discute as limitações da experiência; a Seção 6 apresenta os trabalhos relacionados e, por fim, a Seção 7 apresenta as considerações finais e trabalhos futuros.

\section{Caso de Observação: Sistema WEB para Saúde}

O sistema de software WEB foi idealizado para substituir o processo manual de registro de informações clínicas, que antes era realizado por meio de documentos físicos usados para anotações sobre consultas e evolução do tratamento do tabagismo no IDT/UFRJ. 

O IDT/UFRJ tem como missão promover o contínuo aperfeiçoamento dos sistemas de saúde. Entre suas diversas áreas, destaca-se o trabalho relacionado ao antitabagismo, que abrange tanto o cigarro convencional quanto o VAPE, com foco no acompanhamento de pacientes fumantes que buscam melhorar sua qualidade de vida  ao deixar o hábito de fumar. Para isso, é realizado um acompanhamento clínico periódico, durante o qual são coletados marcadores individuais e realizada a anamnese do paciente, com o objetivo de atualizar o prontuário médico (armazenado em fichas individuais) e obter dados que permitam observar a evolução do tratamento.

Tem sido observado que o sucesso do tratamento está relacionado à frequência das visitas dos pacientes ao consultório e às motivações oferecidas pela equipe médica. No entanto, manter o contato constante com cada paciente é um desafio, motivo pelo qual o IDT/UFRJ decidiu construir um aplicativo móvel para \textit{Android} com o objetivo de apoiar o fumante a abandonar definitivamente o cigarro. Esse aplicativo está em desenvolvimento, de forma convencional, em nossa organização, e permite coletar informações relevantes para o acompanhamento clínico individual. 

Entretanto, as informações coletadas pelo aplicativo precisam ser transmitidas e armazenadas em uma base de dados localizada no IDT/UFRJ, onde poderão ser consultadas e utilizadas pela equipe de saúde. No entanto, ainda não existia uma solução de software WEB com uma base de dados integrada capaz de realizar a gestão do recebimento, armazenamento e consulta desses dados, por meio de \textit{dashboards}  e outras funcionalidades de apoio à decisão clínica, o que motivou a proposta de desenvolvimento da versão WEB do sistema. Um sistema desse tipo, voltado ao acompanhamento de pacientes fumantes no contexto da pneumologia, mostra-se extremamente útil para organizar os atendimentos, monitorar a evolução clínica e viabilizar intervenções personalizadas, sobretudo para aqueles em acompanhamento no IDT/UFRJ.

Neste contexto, e considerando que o aplicativo \textit{Android} estava em construção e com previsão de entrega para o início do segundo semestre de 2025, o IDT/UFRJ vislumbrou, em meados de abril do mesmo ano, a possibilidade de construir também a parte WEB do sistema. Essa nova etapa surgiu a partir da necessidade de contemplar cenários de uso ainda não identificados, projetados ou implementados. O grande problema, no entanto, é que o IDT/UFRJ, como toda organização pública, enfrenta limitações de tempo e recursos: suas equipes de saúde estão sobrecarregadas e precisam de acesso a uma solução de software funcional assim que o aplicativo móvel esteja disponível. Diante disso, nosso desafio passou a ser encontrar meios e instrumentos de engenharia que viabilizassem a construção de uma versão inicial, operacional e com qualidade da aplicação WEB, mesmo diante da limitação de disponibilidade contínua dos stakeholders envolvidos. 

Participaram do projeto duas médicas do IDT/UFRJ, especialistas no domínio do problema que atuaram como principais stakeholders externos, e 15 engenheiros de software considerados stakeholders internos, todos com no mínimo um ano de experiência na indústria (ver Tabela \ref{tab:time_desenvolvimento}). Desses engenheiros, dois não atuam diretamente em pesquisa. Adicionalmente, o perfil de formação do time pode ser descrito com um engenheiro de software e pesquisador master (líder do projeto), cinco doutorandos, sete mestrandos e os outros dois são estudantes no último ano da graduação em Engenharia de Computação e Informação. Os engenheiros de software possuíam familiaridade com ferramentas de IA generativa e com atividades tradicionais do ciclo de vida do software. No entanto, até então, nenhum deles havia utilizado ferramentas de IA generativa de forma integrada a todo o  processo de desenvolvimento de software antes deste caso.

\begin{table}[htbp]
\centering\scriptsize
\caption{Perfil Básico do Time de Desenvolvimento}
\label{tab:time_desenvolvimento}
\begin{tabular}{|p{1.8cm}|p{3.1cm}|p{2.4cm}|}
\hline 
\textbf{Desenvolvedor} & \textbf{Área de Atuação} & \textbf{Nível Profissional} \\
\hline
Dev1 & Engenharia de Software & Master (Líder)\\ 
\hline
Dev2 & Engenharia de Requisitos & Pleno\\ 
\hline
Dev3 & Engenharia de Requisitos & Pleno\\ 
\hline
Dev4 & Especificação de Requisitos & Pleno\\ 
\hline
Dev5 & Especificação de Requisitos & Pleno\\ 
\hline
Dev6 & Projeto & Pleno\\ 
\hline
Dev7 & Projeto & Pleno\\ 
\hline
Dev8 & Gerência de Projetos & Sênior\\ 
\hline
Dev9 & Gerência de Projetos & Júnior\\ 
\hline
Dev10 & Codificação & Sênior\\ 
\hline
Dev11 & Codificação & Pleno\\ 
\hline
Dev12 & Qualidade de Software & Sênior\\ 
\hline
Dev13 & Qualidade de Software & Sênior\\ 
\hline
Dev14 & Qualidade de Software & Pleno\\ 
\hline
Dev15 & Projeto & Júnior\\ 
\hline
\end{tabular}
\end{table}

Diante disso, antes do início das atividades práticas, os participantes receberam um treinamento introdutório sobre conceitos relacionados a IA generativa aplicados à ES. Considerando que equipes de desenvolvimento de software, em sua maioria, não recebem treinamento formal sobre o uso de LLMs e possuem apenas conhecimento básico sobre eles \cite{falcao2024investigating}. Após isso, os desenvolvedores foram organizados em diferentes times, de acordo com sua familiaridade com as fases de desenvolvimento. Também foram apresentados ao contexto clínico do projeto, aos fluxos atuais de atendimento a pacientes em tratamento de cessação do tabagismo e problemas enfrentados com o uso de registros manuais.

A solução de software WEB deve contemplar três perfis de usuário: (i) profissionais de saúde, responsáveis por cadastrar, visualizar e atualizar os dados clínicos e comportamentais dos pacientes, além de registrar observações a cada visita realizada; (ii) residentes, com acesso restrito à visualização dos dados dos pacientes; e (iii) administradores, com acesso total ao sistema, incluindo o gerenciamento de usuários e permissões. O sistema também deve prever a integração com o aplicativo móvel utilizado pelos próprios pacientes, por meio do qual são coletados dados comportamentais para o tratamento, como recaídas, sintomas de abstinência e histórico de consumo de cigarros.

\section{O Processo Construtivo com IA Generativa}

\begin{figure*}
    \Description { Processo }
    \centering
    \includegraphics[width=1\linewidth]{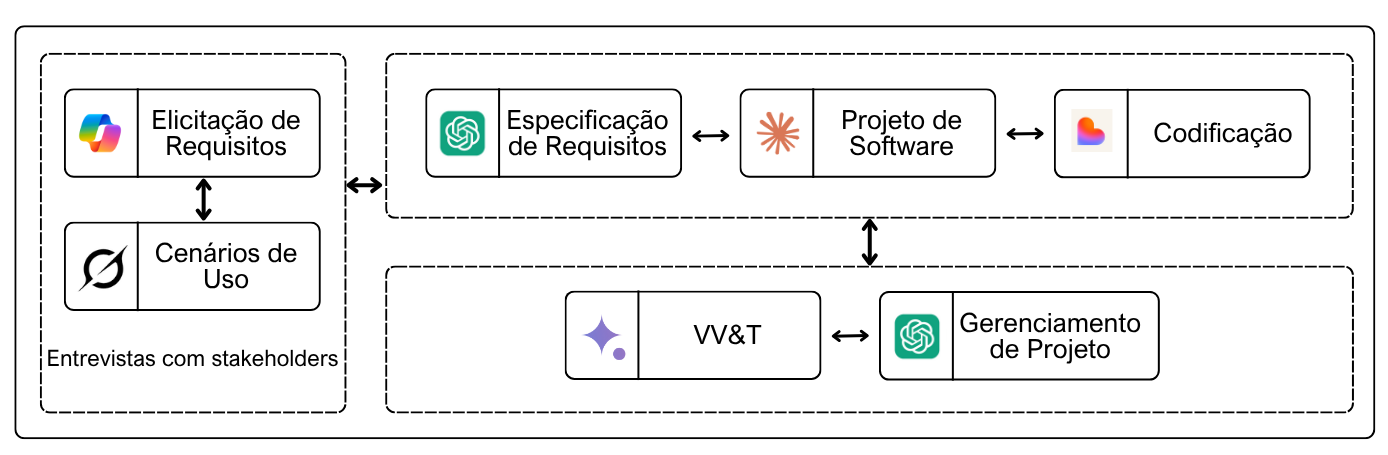}
    \caption{Processo Construtivo com IA}
    \label{fig:processo}
\end{figure*}

A fim de observar a influência das tecnologias de IA generativa, decidimos organizar o processo de construção considerando as etapas clássicas da engenharia de software: elicitação de requisitos, coleta de cenários de uso, especificação de requisitos, projeto, codificação, testes e gerenciamento, conforme pode ser visto na Figura \ref{fig:processo}. Embora representadas de forma sequencial, essas etapas ocorreram iterativamente, como estratégia de investigação e não como um modelo de cascata. Essa decisão foi motivada pela necessidade de observar de forma sistemática o papel da IA generativa em cada atividade, garantindo a rastreabilidade entre os artefatos.

O desenvolvimento teve início com a elicitação de requisitos por meio de entrevistas com a participação das especialistas no domínio, seguida pelos cenários de uso e progressão pelas demais fases, frequentemente de forma sobreposta. Por exemplo, os planos de testes funcionais foram planejados assim que os cenários de uso e os requisitos ficaram disponíveis, sendo executados após a disponibilização de uma versão utilizável do sistema. A etapa de manutenção não foi contemplada por envolver atividades posteriores à entrega e uso efetivo do produto pelos usuários finais.

Para atender esse processo, conforme mencionado, a equipe de desenvolvimento foi organizada em sete times de trabalho: elicitação de requisitos (Dev1, Dev2, Dev3 e stakeholders), cenários de uso (Dev1 e stakeholders), especificação de requisitos (Dev4 e Dev5), projeto do software (Dev6, Dev7 e Dev15), gerenciamento do projeto (Dev8 e Dev9), codificação (Dev10 e Dev11) e verificação, validação e testes (VV\&T) (Dev12, Dev13 e Dev14). 

Cada time foi responsável por conduzir as atividades da sua respectiva fase, o que incluía a criação de \textit{prompts}, a interação com ferramentas de IA generativa e a geração dos artefatos. Os times de gerenciamento de projeto e VV\&T trabalharam de forma transversal às outras atividades. Esses artefatos foram avaliados e validados por desenvolvedores de outros times, conforme representado na Figura \ref{fig:processo}. Além disso, a Tabela \ref{tab:ferramentas} apresenta uma visão geral das ferramentas de IA generativa, bem como dados de entrada (insumo para os modelos) e saídas de cada fase. 

\begin{table*}[htbp]
\centering\scriptsize
\caption{Artefatos Gerados ao Longo do Processo Construtivo com IA}
\label{tab:ferramentas}
\begin{tabular}{|p{2cm}|p{3cm}|p{3cm}|p{3cm}|p{4.5cm}|}
\hline 
\textbf{Fase} & \textbf{Entrada} & \textbf{Ferramenta ou Modelo Utilizado} & \textbf{Saídas} & \textbf{Papel da IA}\\
\hline
Elicitação de Requisitos & Entrevistas com stakeholders & MS CoPilot (\url{https://copilot.microsoft.com/}) & Documento de visão & Para aprimorar a clareza e a articulação textual. \\ 
\hline

Casos de Uso & Documento de requisitos, formulário de história clínica & ChatGPT-4.1 (\url{https://chatgpt.com/}) & Documento de Casos de uso com descrição de fluxos principal, alternativo e de exceção & Para gerar fluxos iniciais (principal, alternativo, exceção).
\\ 
\hline

Especificação de Requisitos & Documento de visão, cenários de uso, formulário de história clínica & ChatGPT-4o (\url{https://chatgpt.com/}) & Documento com 41 requisitos, incluindo funcionais (28) e não funcionais (13) & Para geração inicial dos requisitos, bem como a classificação e priorização.
\\ 
\hline

Design de Projeto & Documento de requisitos, formulário de história clínica & Claude-3.5 (\url{https://claude.ai/}) & Diagrama de Arquitetura, Modelo de Dados & Para apoiar a definição da arquitetura do sistema, componentes e fluxos de dados.
\\ 
\hline

Gerenciamento do Projeto & Documento de visão e cenários de uso & ChatGPT-o4-mini (\url{https://chatgpt.com/}) & Plano de Projeto (modelo clássico) & Para elaborar um plano de projeto, considerando prazos e intenções gerais do uso de ferramentas de IA generativa.
\\ 
\hline

Cenários de Uso & Documento de visão & Grok (4.0) (\url{https://grok.com/}) & Documento com oito cenários de uso & Para gerar cenários de uso.
\\ 
\hline

Codificação & Documento de requisitos, fomulário de história clínica, casos de uso & Lovable \url{https://lovable.dev} & Backend, frontend & Para gerar código para o frontend e backend.
\\  
\hline

Verificação, Validação e Testes & Documento de requisitos, fomulário de história clínica, casos de uso & Gemini 2.5 Flash \url{https://gemini.google.com/} & Documento com Plano de Testes e sugestão de casos de teste & Para gerar planos de testes e associações entre casos de uso e requisitos.
\\ 
\hline

\end{tabular}
\end{table*}

Para interagir com as ferramentas de IA, os \textit{prompts} foram estruturados com base nas diretrizes da \textit{Google}\footnote{Prompting Guide 101: \url{https://services.google.com/fh/files/misc/gemini-for-google-workspace-prompting-guide-101.pdf}} para uso de ferramentas de IA generativa. Esses \textit{prompts} foram organizados em cinco componentes principais: (i) \textit{persona}, que define o papel do modelo como especialista em requisitos de software; (ii) \textit{tarefa}, que descreve a atividade a ser realizada, como a especificação, classificação e priorização de requisitos; (iii) \textit{contexto}, com a descrição do sistema e os documentos produzidos nas fases iniciais; (iv) \textit{exemplo}, apresentando modelos de requisitos funcionais e não funcionais; e (v) \textit{formato de saída}, que especifica como o conteúdo gerado deve ser estruturado. É importante destacar que cada artefato produzido no projeto precisou ser convertido para o formato \textit{Markdown} (.md) a fim de ser utilizado como insumo nos modelos de IA. A Tabela \ref{tab:PROMPT} apresenta um extrato do \textit{prompt} utilizado na etapa de especificação de requisitos. As próximas subseções detalham cada fase do processo e as ferramentas utilizadas.

\begin{table*}[h]
\centering\small
\caption{Extrato de \textit{Prompt} Usado na Especificação de Requisitos}
\label{tab:PROMPT}
\begin{tabular}{|p{17cm}|}
\hline 
\textbf{\# Persona} \\
\hline
Você é uma pessoa engenheira de requisitos de software sênior. \\ \hline
\textbf{\# Tarefa} \\
\hline
Você recebeu a demanda de construir os requisitos de software funcionais e não funcionais a partir do contexto fornecido a seguir. Você também deve priorizar os requisitos dentro das categorias [essencial, importante, desejável]. \\ 
\hline
\textbf{\# Contexto} \\
\hline
O Instituto de Doenças do Tórax (IDT) da Universidade Federal do Rio de Janeiro (UFRJ) tem em sua missão o objetivo de realizar o contínuo aperfeiçoamento dos sistemas de saúde. Dentre as inúmeras frentes de investigação e evolução do conhecimento, se destaca a frente de trabalho antitabagismo (cigarro convencional e VAPE), com acompanhamento de pacientes fumantes que buscam melhorar sua qualidade de vida parando de fumar.
Para isso, é realizado um acompanhamento clínico periódico dos pacientes, no qual se coletam marcadores individuais e é realizada a anamnese do paciente, visando a atualizar o prontuário médico (armazenado em fichas individuais) e obter dados que permitam observar a evolução do tratamento.
Estas informações precisam ser transmitidas e armazenadas em uma base de dados localizada no IDT/UFRJ, onde podem ser consultadas e utilizadas.\\
O documento de proposta apresenta diferentes cenários. O foco nesse momento é no cenário 1:\\
\textbf{1. Cadastro e Perfil do Paciente}\\
- Um novo paciente fumante chega ao consultório. O pneumologista precisa registrar informações detalhadas para acompanhamento.\\
\textbf{Uso do sistema:}\\
- Cadastro de dados pessoais (nome, idade, contato, etc.).\\
- Registro do histórico de tabagismo: quantidade de cigarros por dia, anos de fumo, cálculo de maços-ano.\\\

No entanto, os outros cenários também devem ser levados em consideração, pensando na evolução do sistema e dependência entre requisitos.
Em resumo, o objetivo do sistema é ser uma sistema WEB que permitirá os profissionais de saúde cadastrarem e acompanharem seus pacientes. Além disso, o sistema deve permitir que os profissionais de saúde acompanhem os pacientes que possuem um cadastro no aplicativo associado. \\  \hline

\textbf{\# Exemplos} \\ \hline
Seguem alguns exemplos preliminares de como os requisitos funcionais devem ser descritos: \\

- O sistema deve registrar um novo paciente através de um formulário, como descrito no arquivo \textit{Markdown}.\\
- O sistema deve permitir o registro de visitas de pacientes já cadastrados.\\
- O sistema deve apresentar a listagem dos pacientes cadastrados, apresentando seu nome e situação. Seguem alguns exemplos preliminares de como os requisitos não funcionais devem ser descritos:\\
- O sistema deve ser adaptável a diferentes tamanhos de tela\\
- O sistema deve ser compatível com os navegadores disponíveis no mercado\\ \hline

\textbf{\# Formato Output} \\ \hline
Quero que você gere uma tabela para requisitos funcionais (RF) e outra para requisitos não funcionais (RNF) na seguinte forma: \\
ID Requisito | Descrição | Prioridade\\
RF 01 | O sistema deve... | Essencial \\
\hline
\end{tabular}
\end{table*}

Adicionalmente, estão disponíveis os artefatos produzidos e validados ao longo do projeto\footnote{Artefatos disponíveis em: \url{https://doi.org/10.5281/zenodo.16366063}}, incluindo: (i) os \textit{prompts} estruturados para as interações com ferramentas de IA generativa; (ii) os documentos utilizados como insumos para as ferramentas de IA generativa; e (iii) os artefatos gerados por essas ferramentas. Ressalta-se, contudo, que alguns documentos — como o formulário de história clínica utilizado no IDT/UFRJ e o documento de requisitos — não foram disponibilizados por conterem informações exclusivas e sensíveis do projeto de software desenvolvido.

\subsection{Elicitação de Requisitos e Cenários de Uso}

O processo começou com a fase de elicitação de requisitos, visando entender as necessidades, expectativas e restrições do sistema. O desenvolvedor Dev1 entrevistou duas médicas especialistas e mapeou os fluxos atuais de gestão clínica para identificar problemas, limitações e melhorias possíveis.

A partir das entrevistas, foi elaborada a versão inicial do documento de visão. O Dev1 usou o Copilot (versão gratuita) no Windows 11 para aprimorar o texto, beneficiando-se tanto da facilidade de acesso quanto da capacidade de aprimorar a clareza e a articulação textual \cite{vasilescu2024improving}. Em seguida, o documento foi submetido ao Grok, que gerou sete cenários de uso com base em prompts e no conteúdo do texto. O Dev1 integrou os cenários ao documento de visão e o submeteu à avaliação dos stakeholders (Dev3 e Dev4), que aprovaram todos os cenários e sugeriram a inclusão de dois novos, específicos do IDT/UFRJ, além do formulário de história clínica do paciente. Também indicaram ajustes de estilo, que foram aplicados. Após nova revisão por Dev2 e Dev3, validando a coerência com o domínio, o documento foi apresentado à equipe e passou a orientar as etapas seguintes do projeto. Todo o processo durou cerca de duas semanas e foi elogiado pelos stakeholders externos pela clareza e agilidade na descrição do problema.


\subsection{Especificação de Requisitos}

\begin{figure*}[h]
    \Description { Requisitos}
    \centering
    \includegraphics[width=1\linewidth]{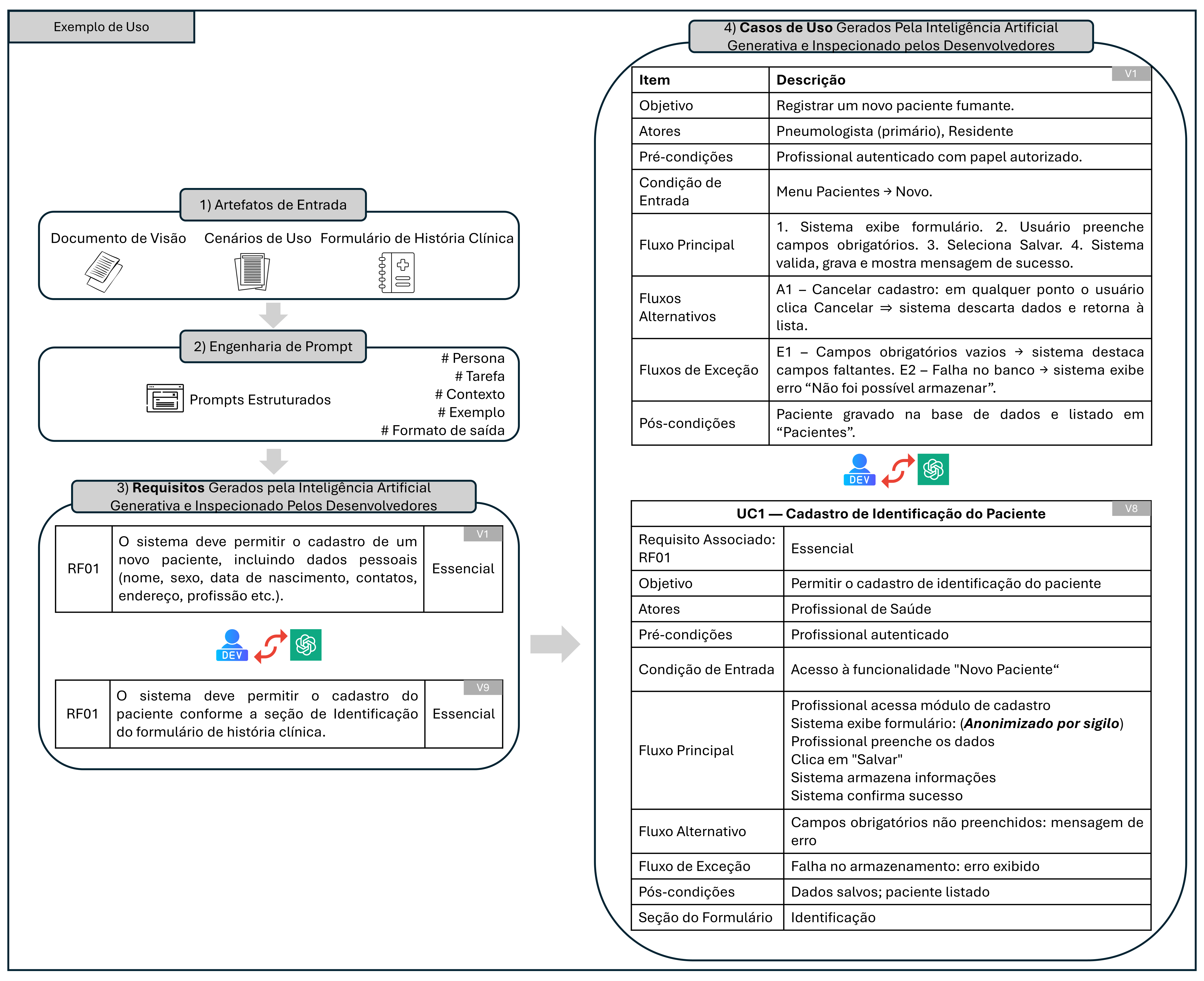}
    \caption{Exemplo do Processo de Especificação de Requisitos e Casos de Uso}
   \label{fig:caso uso}
\end{figure*}

A especificação de requisitos foi feita com apoio do modelo ChatGPT (versão gratuita), escolhido pela facilidade de acesso e familiaridade da equipe com a ferramenta e por estudos anteriores destacarem seu potencial para transformar requisitos de software \cite{marques2024using}\cite{binder2024framework}. A ferramenta ajudou na definição, classificação e priorização de requisitos. As principais fontes de informação foram: o documento de visão, os cenários de uso e um formulário de história clínica utilizado pelas médicas para registrar informações dos pacientes em tratamento, conforme ilustra a Figura \ref{fig:caso uso}. 

A Figura \ref{fig:caso uso} ilustra um exemplo do processo de geração de requisitos e casos de uso. Este processo ocorreu de forma iterativa, resultando em nove versões do documento de requisitos. Cada versão foi inspecionada pelos desenvolvedores Dev1, Dev2 e Dev3, por meio de inspeção \textit{ad-hoc} que identificou omissões, repetições e ambiguidades. As observações levantadas nortearam novas interações com o modelo de IA generativa, permitindo o refinamento progressivo do conteúdo. A versão final foi revisada por seis desenvolvedores (Dev1, Dev2, Dev3, Dev10, Dev14 e Dev15), que sugeriram ajustes pontuais, incorporados antes da finalização do documento.

Com os requisitos consolidados, foram elaborados os casos de uso, contemplando fluxos principal, alternativo e de exceção. Inicialmente, o modelo ChatGPT gerava um caso de uso por requisito, o que resultava em artefatos fragmentados. Para contornar esse problema, foi incluído um exemplo no \textit{prompt}, o que levou à produção de casos mais coesos, funcionais e alinhados ao domínio. Todo o processo levou cerca de duas semanas e resultou em sete versões de caso de uso. 

\subsection{Projeto do Software}

A fase de projeto do software definiu a arquitetura do sistema, componentes e fluxos de dados. Foi conduzida por Dev6 e Dev7, com apoio do modelo Claude Opus 4 (versão gratuita). O \textit{prompt} utilizado foi baseado no documento de requisitos e no formulário de história clínica. O modelo Claude.AI foi escolhido por sua capacidade técnica de processar texto e imagens simultaneamente, o que o tornou adequado para tarefas que envolvem esquemas visuais e descrições técnicas detalhadas.

A primeira versão dos modelos de projeto apresentou falhas, como redundâncias e ambiguidades. Após uma revisão manual, foi gerada uma segunda versão mais coesa por meio de nova interação com a IA. Essa versão passou por uma revisão colaborativa com Dev1, Dev10 e Dev11, que identificaram novas inconsistências. Após um terceiro refinamento do \textit{prompt}, obteve-se um modelo de projeto consolidado e alinhado aos objetivos gerais do sistema. Todo o processo durou cerca de uma semana.

\subsection{Gerenciamento do Projeto}

Esta fase buscou estabelecer os processos construtivos que deveriam ser seguidos. O primeiro modelo para gestão do projeto foi elaborado com base no documento de visão e cenários de uso iniciais. Para isso, o time de gerenciamento de projeto utilizou o ChatGPT para geração do plano de projeto. Um \textit{prompt} foi preparado e os prazos e intenções gerais do uso de tecnologias com IA generativa foram devidamente informados. Entretanto, o plano gerado não conseguiu contemplar as expectativas de projeto, apresentando um conjunto de atividades alinhado com os modelos convencionais de desenvolvimento, por exemplo, o MPS.BR\footnote{Guia Geral MPS de Software:2024: https://softex.br/download/guia-geral-mps-de-software2024/}, sem integração dessas novas tecnologias nas atividades de projeto.

Uma segunda versão do plano de projeto foi gerada após a primeira versão do sistema ter sido construída em um ciclo de transformação completo. Mesmo assim, o plano sugerido manteve o roteiro inicial, prevendo um conjunto de atividades e tarefas incompatíveis com as restrições de construção do sistema. Caso esse plano tivesse sido seguido de forma estrita, não apenas teríamos atrasado as entregas, como também poderíamos comprometer a construção efetiva dos modelos de projeto, prejudicando a adaptação contínua necessária em projetos que envolvem o uso de IA generativa. Essa experiência evidenciou que, embora as ferramentas de IA generativa possam auxiliar na elaboração de planos iniciais, elas ainda apresentam limitações importantes na adaptação a contextos específicos e dinâmicos, exigindo supervisão crítica e ajustes manuais por parte da equipe de projeto.

\subsection{Codificação}

A fase de codificação implementou os requisitos, casos de uso e o design do projeto em um sistema WEB  funcional, com desenvolvimento do \textit{frontend}, \textit{backend} e das camadas de integração.

A escolha da ferramenta de apoio foi baseada em um mapeamento prévio de soluções de IA generativa incluindo: v0, Bolt\footnote{Bolt: \url{https://bolt.new/}}, Replit\footnote{Replit: \url{https://replit.com/}} e Lovable. Todas foram avaliadas a partir do mesmo \textit{prompt} e do formulário de história clínica. A ferramenta Lovable foi selecionada por atender melhor aos requisitos definidos e por integrar de forma mais eficiente as diferentes camadas da aplicação.

Lovable é uma ferramenta de desenvolvimento baseada em IA generativa que permite criar aplicações WEB por meio de comandos em linguagem natural. Ela gera automaticamente o código para \textit{frontend} e \textit{backend}, aceita texto e imagens como entrada e oferece edição visual em tempo real. Por ser intuitiva e eficiente, é indicada para desenvolvedores, designers e pessoas sem experiência técnica que desejam criar soluções digitais de forma ágil. No entanto, sua utilização depende da contratação dos serviços. Embora a intenção inicial do projeto fosse utilizar apenas tecnologias de acesso aberto, optamos pela Lovable devido à disponibilidade de créditos fornecidos para os desenvolvedores.

O uso da ferramenta exigiu a criação de artefatos de projeto, como uma base de conhecimento com descrição, objetivos, perfis de usuários, requisitos funcionais e não funcionais (segurança, desempenho e responsividade), além de diretrizes visuais (paleta de cores, tipografia via imagens exemplo). A stack definida inclui \textit{frontend} em \textit{React} com \textit{Tailwind CSS} e \textit{backend} com \textit{Supabase} (PostgreSQL). Em seguida, cada caso de uso especificado foi transformado em um \textit{prompt} individual, estruturado com base nas melhores práticas sugeridas pela própria Lovable\footnote{Lovable - Prompt Engineering: https://docs.lovable.dev/prompting/prompting-one}, conhecidas como \textit{framework} CLEAR (Conciso, Lógico, Explícito, Adaptativo e Reflexivo) \cite{LO2023102720}. 

A implementação seguiu as recomendações da Lovable, começando pelas funcionalidades essenciais, como autenticação e gestão de usuários. Após revisão, foram desenvolvidas outras funções, como cadastro de pacientes, histórico de tabagismo e cálculos automáticos baseados nos formulários e perfis de acesso. A ferramenta gerava propostas alinhadas aos requisitos, mas que exigiam ajustes finos feitos por meio de múltiplas iterações usando o recurso \textit{``Try to fix”}. Para corrigir falhas na exibição ou ajustes na interface, foram usados novos \textit{prompts} curtos.

\subsection{Verificação, Validação e Testes}

O time de verificação, validação e testes escolheu inicialmente o ChatGPT, considerando sua popularidade, acessibilidade e os resultados preliminares do mapeamento de tecnologias de IA realizado. A primeira versão do plano de testes foi gerada com base no documento de visão, nos cenários de uso e no formulário de história clínica do paciente. Para orientar a geração, foi utilizada no \textit{prompt} a persona de um ``Analista de Teste de Software Sênior''.  

A geração inicial dos casos de teste com o ChatGPT foi interrompida devido à falta de detalhamento e às limitações do uso gratuito. A equipe então adotou o Gemini como ferramenta principal. Apesar da reaplicação do \textit{prompt}, os planos de testes continuaram pouco detalhados, exigindo ajustes por meio de novos \textit{prompts}. Com o avanço do projeto, foi necessário integrar versões atualizadas dos documentos, mas o modelo apresentou dificuldades de assimilação, exigindo o reenvio das informações.

Durante a geração, o Gemini inseriu marcadores de citação, que foram removidos a pedido da equipe. Também foram solicitadas a inclusão de um menu com links, o detalhamento de casos de exceção e a correção de perfis, substituindo o perfil `ADMIN' pelos perfis `Profissional da saúde' e `Residente'. Por fim, o plano de testes foi ajustado para garantir que todas as funcionalidades fossem contempladas. Além disso, foi elaborado um template para o plano de testes. A geração da segunda versão do plano exigiu a refatoração do \textit{prompt} de entrada, detalhando e alinhando as instruções às diretrizes de \textit{prompt} da equipe de requisitos, o que garantiu maior coesão ao projeto.

Durante a etapa de VV\&T, o plano de testes apontou associações entre casos de uso e requisitos que não estavam especificados  na documentação original. O time de requisitos revisou essas associações e ajustou o plano de testes de acordo. Após várias interações e refinamentos nos \textit{prompts}, o plano final foi aprovado pelo time de VV\&T, atendendo aos objetivos e aos padrões de qualidade do projeto. A geração dos testes automatizados seguiu uma abordagem exploratória, utilizando ChatGPT e Gemini. Outros modelos, como DeepSeek\footnote{DeepSeek: \url{https://www.deepseek.com/}} e Claude, foram empregados com \textit{prompts} padronizados para infraestrutura e interfaces. Enquanto ChatGPT e DeepSeek apresentaram falhas persistentes, o Claude produziu testes funcionais com mínima necessidade de intervenção humana. A eficácia dessas ferramentas depende diretamente da qualidade dos \textit{prompts} e da supervisão especializada. A experiência demonstrou que as ferramentas de IA generativa, embora sejam valiosas para automatizar e gerar artefatos de testes, ainda requerem ajustes constantes e validação humana cuidadosa.

\section{Lições Aprendidas}

A experiência de uso da IA generativa, especificamente do ChatGPT na versão gratuita, para a \textbf{especificação dos requisitos} do sistema WEB evidenciou que a automatização completa dessas atividades ainda não é viável. Embora a ferramenta tenha oferecido um suporte inicial promissor, foram necessárias quatro rodadas de revisão e ajustes manuais dos artefatos por parte dos desenvolvedores para alcançar um resultado aceitável em termos de qualidade e coerência. Durante esse processo, o documento de requisitos gerado apresentou diversos defeitos, como omissões importantes de requisitos, duplicação de itens semelhantes e inadequação na priorização dos requisitos, comprometendo a clareza e a utilidade do documento para orientar as fases subsequentes do desenvolvimento. Esses problemas demandaram inspeção e refinamentos iterativos para corrigir inconsistências e preencher lacunas. 

No que diz respeito à geração dos casos de uso, identificou-se a necessidade de elaborar um exemplo mais sofisticado e detalhado no \textit{prompt} para garantir que o modelo estabelecesse uma associação coerente e integrada entre requisitos e casos de uso. Isso evitou que a geração resultasse em simples correspondências um a um, sem refletir as interdependências e fluxos reais do sistema. Além disso, foi observada uma limitação técnica na geração dos casos de uso, que frequentemente se apresentava de forma incompleta. Na maioria dos casos, foi preciso submeter pelo menos dois comandos adicionais para obter a geração completa dos casos de uso restantes. Um aspecto relevante constatado foi a maior incidência de erros e o surgimento de ``alucinações'', respostas incorretas ou sem fundamento, nas últimas partes geradas. Esse ponto reforça a necessidade de ``relembrar'' o contexto completo para o modelo de linguagem em cada solicitação, garantindo que ele tenha acesso a todo o histórico e informações necessárias para produzir uma saída consistente e coerente. 

Um desafio adicional com a ``memória'' e histórico das interações com o ChatGPT foi identificado durante o processo de desenvolvimento. No início da geração dos requisitos, o modelo produziu uma especificação incorreta de uma funcionalidade, que foi corrigida manualmente na versão final do documento. Entretanto, durante a geração dos casos de uso, a mesma funcionalidade foi descrita com a mesma incorreção. Esse episódio evidencia a dificuldade de garantir a correção das respostas geradas pela IA gerativa, mesmo após inspeções manuais detalhadas dos artefatos. 

Dessa forma, verificamos que o uso de IA generativa para a especificação de requisitos pode ser bastante vantajoso, pois facilita e agiliza o processo. No entanto, ainda é imprescindível a supervisão humana constante, com a verificação e validação dos requisitos gerados para garantir sua qualidade. Neste estudo, a aplicação de inspeção de software \textit{ad-hoc} foi fundamental para assegurar a confiabilidade dos artefatos produzidos. Mesmo assim, o gerenciamento do contexto pelo modelo e a elaboração de \textit{prompts} otimizados continuam sendo desafios significativos para a geração de uma especificação precisa e consistente. 

A utilização de IA generativa no \textbf{projeto do software} demonstrou potencial para acelerar a criação inicial de arquiteturas, diagramas e modelos de dados, embora tenha revelado importantes limitações. Um dos principais aprendizados foi que o nível de detalhamento e clareza dos \textit{prompts} influencia diretamente a qualidade dos artefatos gerados. \textit{Prompts} genéricos ou vagos, mesmo quando fundamentados nos documentos de projeto, resultaram em arquiteturas simplificadas, com sobreposição de responsabilidades, fluxos de dados pouco claros e ausência de informações essenciais. Para mitigar essas deficiências, foi necessário investir em descrições detalhadas dos componentes, fluxos de interação e restrições técnicas, além de incluir exemplos explícitos de diagramas e modelos similares validados na elaboração dos \textit{prompts}.

Um desafio adicional foi manter a consistência ao longo de múltiplas iterações, que nem sempre foram incrementais. A cada nova geração de artefatos ou ajustes, a ferramenta de IA generativa frequentemente desconsiderava detalhes previamente estabelecidos ou alterava partes já validadas do projeto. Isso exigiu uma gestão rigorosa do contexto, com rastreamento detalhado das mudanças e supervisão constante, além da manutenção de documentação paralela para versionamento dos resultados obtidos. A experiência demonstrou que a IA generativa tem potencial para ser uma ferramenta valiosa para apoiar o projeto do software, contribuindo tanto para a produtividade quanto para a inspiração criativa. Contudo, seu uso requer \textit{prompts} extremamente bem estruturados, exemplos claros e uma atuação contínua e crítica dos arquitetos humanos. Mais uma vez, a inspeção foi fundamental para garantir a qualidade dos artefatos gerados. Com isso, é interessante observar que os modelos de projeto foram úteis tanto para o entendimento da equipe quanto para a possível solução de software. Entretanto, estes modelos pouco influenciaram a codificação, sugerindo a necessidade de avaliar sua efetiva necessidade em casos semelhantes de projeto.

No \textbf{gerenciamento de projeto}, percebemos que a ferramenta reproduziu um modelo de desenvolvimento tradicional desalinhado com as demandas reais do projeto, especialmente considerando as limitações e dinâmicas de um contexto público e inovador com o uso de tecnologias de IA. Isso demonstrou que não basta automatizar processos antigos: é fundamental adaptar o gerenciamento para refletir as particularidades da aplicação de novas tecnologias, como as baseadas em IA generativa, considerando a necessidade de flexibilidade, ciclos iterativos curtos e menor dependência da presença dos stakeholders. Caso contrário, o planejamento pode se tornar um entrave e impedir o progresso do projeto.

Na fase de \textbf{codificação}, constatou-se um relacionamento intrínseco entre a qualidade dos resultados obtidos nas etapas anteriores de requisitos e casos de uso e o grau de detalhamento das informações fornecidas à ferramenta Lovable. \textit{Prompts} genéricos ou estruturados de forma insuficiente frequentemente levaram a funcionalidades incompletas ou comportamentos inesperados. Nesse contexto, reforçou-se a importância do uso de documentos em formato \textit{Markdown} (.md), que permitiu manter uma estrutura hierárquica clara das informações, permitindo descrever com precisão as tarefas e casos de uso, bem como detalhar explicitamente os campos, tipos de dados e suas obrigatoriedades, quando aplicáveis.

A inclusão de imagens como referência nos \textit{prompts} foi essencial para garantir consistência visual nas interfaces, reduzindo alterações inesperadas nas iterações. A estratégia de dividir a construção por casos de uso também se mostrou eficaz para modular o desenvolvimento. Contudo, funcionalidades mais complexas como o cálculo automático de pontuação, o controle de permissões por perfil de usuário e a atualização dinâmica da interface não foram corretamente implementadas em uma única iteração. Essas limitações evidenciaram a dificuldade da ferramenta em interpretar regras lógicas complexas, manter estado entre componentes e garantir consistência ao longo do tempo. Foi necessário recorrer a múltiplas interações e ajustes pontuais para atingir os resultados esperados. 

Observou-se ainda que, à medida que o sistema evoluía, tornava-se mais difícil aplicar alterações específicas via \textit{prompt}, visto que intervenções não solicitadas passaram a ocorrer com maior frequência. Assim, entendemos que o uso de IA generativa na construção de sistemas de software requer delimitação clara de escopo, modularidade no desenvolvimento, criação de novos artefatos, e atenção especial às limitações da ferramenta diante de regras de negócio complexas e interfaces interativas.

Visando contornar tais limitações e aprimorar a eficiência do processo construtivo com tecnologias de IA generativa, sugerimos a integração da ferramenta Lovable (ou outra plataforma similar baseada em agentes) com ferramentas complementares. Observamos que uma abordagem viável é combinar a geração inicial da aplicação pela ferramenta com correções pontuais e ajustes mais precisos realizados por meio de LLMs de uso geral como o ChatGPT. Também é recomendável considerar a exportação do código gerado para editores avançados, como o Cursor\footnote{Cursor: \url{https://cursor.com/}}, com integração direta ao \textit{Github}\footnote{GitHub: \url{https://github.com/}}, permitindo edição assistida por IA generativa, controle de versões mais robusto e colaboração mais efetiva na resolução de problemas. Vale ressaltar que tal abordagem necessita de uma maior experiência dos desenvolvedores que precisam identificar no código os componentes que necessitam de ajustes antes de os submeterem ao assistente ou solicitarem uma correção específica.

A experiência com o uso de ferramentas de IA generativa, como ChatGPT e Gemini, nas atividades de \textbf{garantia da  qualidade} evidenciou o potencial dessas tecnologias para apoiar tarefas específicas da ES. Em particular, destacaram-se como úteis para a geração inicial de artefatos e para acelerar a execução de atividades repetitivas ou operacionais, como a estruturação de cenários de teste ou a identificação de casos básicos a partir de requisitos funcionais simples. Entretanto, a experimentação prática também revelou limitações importantes. Em muitos casos, os planos de teste gerados careciam de profundidade, não contemplando cenários de borda, exceções ou particularidades do domínio da aplicação. Além disso, foi comum a inserção de informações genéricas ou desconectadas do contexto real do projeto, indicando que a IA generativa apresenta dificuldade em correlacionar múltiplas fontes técnicas (como documentos de requisitos, regras de negócio e especificações de sistema) e interpretar adequadamente nuances mais complexas.

Por exemplo, a implantação inicial do sistema WEB também foi realizada por meio do Lovable, o que facilitou o acompanhamento do processo ao disponibilizar hospedagem automática em nuvem e fornecer uma URL pública para acesso da aplicação. Destaca-se a integração nativa da plataforma de desenvolvimento com o Supabase, solução \textit{open source} responsável pela autenticação, armazenamento de dados em PostgreSQL e aplicação de políticas de \textit{Row Level Security} (RLS). Essa integração é feita via \textit{API Key}, o que permite ao Lovable executar operações no Banco de Dados. Contudo, observou-se que o código gerado pelo Lovable incluía a chave de \textit{API} diretamente no código-fonte, prática inadequada e que pode expor informações sensíveis e comprometer a segurança. Esse defeito foi corrigido manualmente, transferindo as credenciais para um arquivo \textit{.env} e realizando a importação adequada no ambiente de execução. 

Além disso, a precisão dos \textit{prompts} se mostrou fundamental para a configuração das regras de acesso para garantir a segurança e consistência da aplicação. \textit{Prompts} mal definidos podem levar a LLM a sugerir a remoção de restrições de autenticação ou enfraquecimento das políticas de RLS, criando riscos para a integridade dos dados. Esse tipo de vulnerabilidade está documentado na \textit{Common Vulnerabilities and Exposures}\footnote{Common Vulnerabilities and Exposures: \url{https://nvd.nist.gov/vuln/detail/CVE-2025-48757}} (CVE-2025-48757) da ferramenta, que detalha este tipo de exposição. Por fim, para viabilizar a execução da aplicação em ambientes locais, foi necessário emitir um \textit{prompt} adicional ao Lovable, solicitando a geração dos arquivos de configuração em \textit{Docker}\footnote{Docker: \url{https://www.docker.com/}}.

Essas limitações exigiram um esforço adicional de inspeção por parte da equipe, especialmente na validação dos critérios de aceitação, completude dos cenários e aderência aos objetivos de teste. Esse processo revelou que o uso da IA, embora promissor, ainda depende fortemente da atuação crítica de profissionais qualificados, capazes de revisar, complementar e adaptar os resultados gerados para garantir a qualidade do software construído. Outro ponto central observado foi a importância da engenharia de \textit{prompt}. Conforme mencionado anteriormente, quanto mais específico, contextualizado e bem estruturado o \textit{prompt}, maior a chance de obter respostas úteis e alinhadas às necessidades do projeto. Isso torna claro que o domínio técnico do profissional não é apenas desejável, mas essencial para transformar a IA generativa em uma aliada real nos processos de qualidade.

Desta forma, observamos que o uso de IA generativa em atividades de garantia da qualidade deve ser encarado como apoio e não substituição profissional. A eficácia dessas ferramentas depende diretamente da engenharia de \textit{prompt}, área ainda incipiente e com pouca informação, e da atuação crítica de profissionais experientes, reforçando a importância de revisões e inspeções manuais como pilares da qualidade de software. Assim como nas empresas pesquisadas, também percebemos que o atendimento a requisitos legais era frequentemente considerado suficiente, o que evidencia a necessidade de tratar a ética como um requisito não funcional e integrá-la ativamente às práticas de desenvolvimento.

Essa experiência evidenciou que questões éticas não podem ser tratadas de forma isolada ou superficial. Princípios como responsabilidade e transparência foram fundamentais para garantir que os artefatos gerados fossem criticamente avaliados, corrigidos e rastreados. Em todas as etapas, desde a elicitação de requisitos até a validação, verificação e testes, foi necessário manter a supervisão ativa de humanos para assegurar que os resultados da IA estivessem de acordo com os objetivos do sistema WEB. Em linha com os achados de Baldassarre \textit{et al.} \cite{baldassarre2024polaris}, aprendemos que a ética deve ser integrada desde o início ao processo de desenvolvimento. Além disso, considerações sobre justiça também se mostraram relevantes: ao elaborar requisitos e casos de uso com apoio da IA generativa, foi necessário garantir que não houvesse vieses ou omissões que pudessem prejudicar perfis específicos de usuários.

\section{Limitações da Experiência Relatada}

Apesar dos esforços para garantir rigor na condução e documentação deste relato de experiência, reconhecemos a existência de algumas limitações que podem influenciar a interpretação dos resultados. 

A avaliação da qualidade dos artefatos gerados pelas ferramentas de IA generativa, por exemplo, envolveu certo grau de subjetividade por parte dos desenvolvedores. Para mitigar esse aspecto, adotamos estratégias como revisões por pares, discussões internas e registro da rastreabilidade dos artefatos, visando maior consistência e transparência nas avaliações. Além disso, o conhecimento prévio dos participantes sobre IA generativa e seu nível de engajamento com as atividades podem ter influenciado diretamente os resultados observados. Portanto, restringe-se a profundidade das análises e a generalização das conclusões em diferentes domínios.

Também reconhecemos que as observações foram realizadas em um único projeto, com uma equipe específica de desenvolvedores e em um contexto particular de desenvolvimento, o que pode limitar a generalização dos achados. Como forma de mitigar essa limitação, descrevemos em detalhes o cenário investigado, as ferramentas utilizadas e os critérios adotados nas decisões, de modo a possibilitar que outros profissionais avaliem a aplicabilidade dos resultados em seus próprios contextos. Reconhecemos também que a divisão da equipe de desenvolvimento e das ferramentas de IA, embora justificáveis no contexto do estudo, pode ter restringido a abrangência das respostas à pergunta de pesquisa e a generalização dos resultados.

Por fim, a rápida evolução das ferramentas de IA generativa representa um desafio adicional, pois algumas observações podem se tornar rapidamente desatualizadas. Para lidar com esse aspecto, registramos de forma explícita as versões das ferramentas utilizadas e os períodos em que foram aplicadas, garantindo clareza quanto ao contexto tecnológico em que as experiências foram conduzidas. Cabe ressaltar que as ferramentas utilizadas estavam em suas versões gratuitas, o que impõe restrições técnicas (limites de contexto, tokens e funcionalidades) que podem ter influenciado os resultados.

Além dessas limitações, é importante considerar aspectos éticos associados ao uso de ferramentas de IA generativa. As saídas produzidas por esses sistemas podem refletir vieses presentes nos dados de treinamento, levantar dúvidas quanto à autoria intelectual dos artefatos gerados e influenciar decisões sem oferecer explicabilidade adequada. Embora este relato não tenha identificado casos críticos, reconhecemos que a ausência de mecanismos sistemáticos (e.g., \textit{checklists} éticos, ferramentas de auditoria) para avaliar tais implicações pode representar uma limitação. Futuras investigações poderiam abordar com maior profundidade esses aspectos, incorporando diretrizes éticas específicas e estratégias de mitigação voltadas ao uso responsável da IA.



\section{Trabalhos Relacionados}

Estudos recentes têm explorado o uso de ferramentas baseadas em IA generativa para apoiar atividades específicas do desenvolvimento de software. O trabalho de Zhao \textit{et. al}\cite{zhao2023reqgen} apresentam a ReqGen, uma abordagem baseada em keywords para a geração automática de requisitos de software, utilizando o modelo UniLM. De forma complementar, Nair e Thushara \cite{nair2025nl2code} desenvolveram o NL2Code, um framework híbrido que combina Processamento de Linguagem Natural (PLN) e Engenharia Dirigida por Modelos (MDE) para traduzir requisitos em linguagem natural e diagramas UML em código executável.

Apesar dos avanços, esses estudos concentram-se no uso de IA generativa em tarefas específicas do ciclo de desenvolvimento e em experiências relacionadas à produtividade de desenvolvedores individuais \cite{ebert2023generative}\cite{coutinho2024role}. Em contraste, o presente estudo adota uma perspectiva mais ampla, aplicando ferramentas de IA generativa em múltiplas atividades: elicitação e especificação de requisitos, projeto, gerenciamento de projetos, codificação, testes, experimentação e, de forma transversal, garantia da qualidade. Assim, nosso trabalho avança em relação às pesquisas anteriores ao demonstrar como a IA generativa pode ser incorporada de forma sistêmica e integrada em um projeto de software, indo além do suporte a atividades isoladas.

\section{Conclusão}\label{conclusion}

A experiência aqui relatada, vivenciada em um contexto real de desenvolvimento de um sistema de software WEB voltado à saúde pública, permitiu refletir sobre o papel emergente de ferramentas de IA generativa em atividades centrais do ciclo de vida do software, incluindo o gerenciamento do projeto, a engenharia de requisitos, o projeto de interface, a construção e a garantia da qualidade.  Em resposta à nossa pergunta orientadora — se seria possível desenvolver um sistema de software com qualidade utilizando apenas ferramentas baseadas em IA generativa ao longo de todo o ciclo de desenvolvimento —, concluímos que a resposta é ainda negativa. Apesar dos ganhos de produtividade e da utilidade das ferramentas em tarefas específicas, a qualidade do sistema só foi alcançada graças à atuação humana contínua, especialmente na formulação de \textit{prompts} eficazes, na inspeção técnica e na validação de requisitos e artefatos gerados. Além disso, o uso destas tecnologias provocou a necessidade de utilização de novos artefatos no projeto (\textit{prompts}, artefatos em \textit{Markdown}), os quais demandaram a inserção de atividades adicionais para garantia da qualidade do software.  Assim, a experiência reforça que a IA generativa pode ser uma aliada no processo de desenvolvimento, mas ainda não substitui o papel crítico da \textbf{experiência humana} na garantia da qualidade do software.

Dentre os principais aprendizados, destacamos que o uso dessas ferramentas pode, de fato, trazer ganhos em produtividade, especialmente nas fases iniciais de construção de artefatos ou protótipos, mesmo considerando o surgimento de novos artefatos e atividades de construção. A capacidade de gerar esboços de testes, interfaces a partir de descrições textuais estruturadas, demonstrou-se útil para acelerar ciclos de iteração e fomentar o alinhamento entre a equipe técnica e partes interessadas. Contudo, o potencial das ferramentas está fortemente condicionado à qualidade da interação humano-IA. A engenharia de \textit{prompt} revelou-se uma competência crítica, e os melhores resultados foram obtidos quando os \textit{prompts} foram derivados de documentos bem estruturados, com detalhamento preciso e linguagem sem ambiguidades. A codificação e o detalhamento em \textit{Markdown}, por exemplo, mostraram-se uma estratégia eficaz para organizar informações e orientar a geração de artefatos relevantes.
Ainda assim, a aplicação prática revelou importantes limitações: as ferramentas demonstraram dificuldade em compreender relações complexas entre documentos, interpretar regras de negócio mais sofisticadas e manter a consistência entre diferentes partes do sistema. Em contextos regulados, como o da saúde, essas lacunas representam riscos concretos e exigem que toda saída gerada seja cuidadosamente inspecionada por especialistas. 

Nesse sentido, a experiência reforça que: (i) a IA generativa deve ser vista como uma \textbf{ferramenta complementar}, e não substitutiva; (ii) \textbf{documentação bem estruturada e expressa em linguagem clara} é condição essencial para uma boa interação com sistemas generativos;  (iii) o papel da \textbf{revisão humana e da inspeção técnica permanece insubstituível no ciclo de desenvolvimento}, especialmente quando se busca qualidade, conformidade e confiança, e (iv) precisamos \textbf{realizar uma avaliação crítica de nossos modelos correntes de processos construtivos de software} considerando os novos artefatos e papéis necessários para realizar a efetiva construção de um produto de software com IA.

Por fim, recomendamos que equipes interessadas em adotar essas tecnologias no contexto da engenharia de software o façam de forma incremental, com processos de validação sistemáticos, e reconhecendo que o verdadeiro ganho da IA generativa ocorre quando combinada à expertise de profissionais que compreendem o domínio, os objetivos do sistema e os limites das ferramentas. Uma atenção adicional deve ser dada ao plano de projeto. As estruturas clássicas de organização de processos de desenvolvimento usualmente apresentadas nos modelos de maturidade correntes devem ser evoluídas para contemplar a inserção destas novas tecnologias. 

Como trabalhos futuros, pretendemos aprofundar a integração das ferramentas de IA generativa no ciclo de desenvolvimento, especialmente focando em automação com validação humana. Também buscamos investigar melhores práticas para a engenharia de \textit{prompts}, comparar artefatos produzidos com e sem IA, e avaliar o uso dessas tecnologias em equipes de diferentes tamanhos e perfis. Pretendemos também explorar a mesma abordagem em processos ágeis, para avaliar se os resultados se mantêm nesse contexto. Além disso, queremos explorar o apoio da IA generativa na engenharia de requisitos em nossos projetos de software, ampliando as evidências sobre sua aplicação prática na engenharia de software.

\section*{Artefatos}

O material utilizado no projeto está disponível em: \url{https://doi.org/10.5281/zenodo.16366063}.

\section*{Agradecimentos}

Os autores agradecem intensamente ao Instituto de Doenças do Tórax da Universidade Federal do Rio de Janeiro, representado pelas médicas Michelle Cailleaux e Natália Blanco, por confiarem à nossa equipe a construção do software. Esta experiência é apoiada pelas agências CBV e FBG. Reconhecemos o uso de ferramentas Copilot e ChatGPT para apoiar o refinamento do nosso texto original para melhorar a clareza. Revisamos todos os resultados para garantir o alinhamento com nossa intenção original. O Prof. Travassos é pesquisador CNPq e CNE FAPERJ. CAPES, FAPERJ e CNPq financiam parcialmente este trabalho.

\bibliographystyle{ACM-Reference-Format}
\bibliography{sample-base}

\footnote{ This paper has been accepted in Portuguese and translated to English for publication in the 2025 Brazilian Quality Software Symposium (SBQS), Available at the \textbf{S}BC\textbf{O}PEN\textbf{L}IB - https://sol.sbc.org.br/index.php/indice. Please cite as indicated by SOL.}

\end{document}